\documentclass[superscriptaddress,twocolumn,floatfix,nofootinbib]{revtex4-2}
\usepackage{times,fancyhdr}
\usepackage[dvips]{graphicx}
\usepackage{amsmath,amssymb,bm,dsfont}
\usepackage{color,soul}
\usepackage{mathrsfs}
\usepackage{setspace}
\usepackage{hyperref}

\newtheorem{axiom}{Axiom}
\newtheorem{requirement}{Requirement}
\newtheorem{soln}{Solution}

\begin{document}
\title{What does it take to solve the measurement problem?}
 
\author{Jonte R.\ Hance}
\email{jonte.hance@bristol.ac.uk}
\affiliation{Quantum Engineering Technology Laboratories, Department of Electrical and Electronic Engineering, University of Bristol, Woodland Road, Bristol, BS8 1US, UK}
\author{Sabine\ Hossenfelder}
\affiliation{Frankfurt Institute for Advanced Studies, Ruth-Moufang-Str. 1, D-60438 Frankfurt am Main, Germany}
\date{\today}

\begin{abstract}
We summarise different aspects of the measurement problem in quantum mechanics. We argue that it is a real problem which requires a solution, and identify the properties a theory needs to solve the problem. We show that no current interpretation of quantum mechanics solves the problem, and that, being interpretations rather than extensions of quantum mechanics, they cannot solve it. Finally, we speculate what a solution of the measurement problem might be good for.
\end{abstract}

\maketitle

\section{Introduction}

Quantum mechanics, in its standard formulation (often referred to as the Copenhagen Interpretation), has two different axioms for its time evolution. The one is the deterministic, linear Schr\"odinger equation, the other the non-deterministic, non-linear, and generically non-local collapse of the wave function. The latter is sometimes also referred to as the ``reduction'' or ``update'' of the wave function. 

The collapse of the wave function must be mathematically applied in the event of a measurement, yet the theory leaves unspecified just what constitutes a measurement. While this problem is a century old, it is still hotly debated \cite{Mermin2022NoProblem}. 
We will argue here that this is more than just unsatisfactory, but is a severe shortcoming that requires a solution. Previous accounts of some of the aspects discussed here can be found in \cite{Maudlin1995ThreeMP,leggett2005measprob,weinberg2016}.

Throughout this paper we use natural units ($\hbar=c=1$).

\section{The Axioms}
\label{sec:ax}

Below, we will take an instrumental perspective on quantum mechanics. For the purposes of this paper, quantum mechanics is a mathematical machine. Into this machine we insert some known properties of a system that we have prepared in the laboratory. Then we do the maths, get out a prediction for measurement outcomes, and compare the prediction to the observation. 

The axiomatic framework of this mathematical machine can roughly be summarised as:
\begin{axiom}\label{A1}
    The state of a system is described by a vector $|\Psi\rangle$ in a Hilbert space $\mathscr{H}$.
\end{axiom}

\begin{axiom}\label{A2}
    Observables are described by Hermitian operators $\hat O$. Possible measurement outcomes correspond to one of the mutually orthogonal eigenvectors of the measurement observable $|O_I\rangle$.
\end{axiom}
    
\begin{axiom}\label{A3}
    In the absence of a measurement, the time-evolution of the state is determined by the Schr\"odinger equation $i \partial_t |\Psi \rangle = \hat H |\Psi \rangle$.
\end{axiom}

\begin{axiom}[Collapse Postulate]\label{A4Collapse}
    In the event of a measurement, the state of the system is updated to the eigenvector that corresponds to the measurement outcome $|\Psi\rangle \to |O_I\rangle$.
\end{axiom}

\begin{axiom}[Born's Rule]\label{A5Born}
    The probability of obtaining outcome $|O_I\rangle$ is given by $|\langle \Psi | O_I \rangle|^2$. 
\end{axiom} 

\begin{axiom}\label{A6}
    The state of a composite system is described by a vector $|\Psi\rangle$ in the tensor product of the Hilbert-spaces of the individual systems.
\end{axiom}

This brief summary doesn't do justice to all the subtleties of quantum mechanics. Among other things, it doesn't specify what the Hamiltonian operator is or how one gets the operators corresponding to the measurement observables. However, the question of what those operators look like will not concern us in the rest of this paper. 

Of course, there are many different ways to approach quantum mechanics axiomatically. A notable attempt is for example that proposed by Hardy \cite{hardy2001quantum}. We will here use the above set of axioms because it is the way quantum mechanics is typically taught to students, and we believe that this familiarity will make our argument more accessible. 

It is sometimes questioned whether the Collapse Postulate is actually necessary (e.g. in \cite{zurek2018quantum}). Without it, quantum mechanics would still correctly predict average values for large numbers of repetitions of the same experiment. This is the statistical interpretation suggested by Ballentine \cite{ballentine1970statistical}. 

However, we do not merely observe averages of many experiments: we also observe the outcomes of individual experiments. And we know from observations that the outcome of an experiment is never a superposition of detector eigenstates, nor is it ever a mixed state (whatever that would look like)---a detector either detects a particle or it doesn't, but not both. As Maudlin put it \cite{Maudlin1995ThreeMP}, ``it is a plain \emph{physical fact} that some individual cats are alive and some dead'' (emphasis original). Without the Collapse Postulate, the mathematical machinery of quantum mechanics just does not describe this aspect of physical reality correctly.

This means quantum mechanics without the Collapse Postulate is not wrong, but it describes less of what we observe. The Collapse Postulate is hence useful, and part of the axioms because it increases the explanatory power of the theory. It cannot simply be discarded.

The necessity of the Collapse Postulate to describe observations is not a problem in itself, but it gives rise to the problems discussed below. 

\section{The Problems}
\label{sec:prob}
 
\subsection{The Heisenberg Cut}
\label{ssec:Heisenberg}
 
The most obvious problem with the axioms of quantum mechanics is the term ``measurement,'' which remains undefined. 
 
The need to refer to a measurement strikes one as suspect right away, because quantum mechanics is commonly believed to be a fundamental theory for the microscopic constituents of matter. But if quantum mechanics was fundamental, then the behavior of macroscopic objects (like measurement devices) should be derivable from it. The theory should explain what a measurement is, rather than require it in one of its axioms. 
 
This problem has been known since the earliest days of quantum mechanics and is often referred to as the ``Heisenberg Cut'', alluding to the question just where to make the ``cut'' between the unitary Schr\"odinger evolution and the non-unitary measurement update \cite{Heisenberg1952Cut}.
 
One may object that it is a rather inconsequential problem, because in practice we know that, roughly speaking, measurements are caused by large things. This is how we have used the axioms of quantum mechanics so far, and it has worked reasonably well. Us not knowing just how large a device needs to be to induce a measurement hasn't really been an issue. However, the smaller the measurement devices we can manufacture become, the more pressing the question becomes.
 
That we cannot answer this question has practical consequences already. A few years ago, Frauchinger and Renner argued that quantum mechanics cannot consistently describe the use of itself \cite{frauchiger2018quantum}. But as was pointed out in \cite{relano2018decoherence,zukowski2021physics}, the origin of the inconsistency is that Frauchinger and Renner did not specify what a measurement device is. They treated a measurement merely as a sufficiently strong correlation, which leads to a basis ambiguity that allows mutually contradictory results. The problem was directly created by them not making a Heisenberg Cut.

This alleged inconsistency was later experimentally tested with a setup that, stunningly enough, used single photons as stand-ins for observers that supposedly make measurements \cite{proietti2019experimental}. Now, it may be a matter of debate just exactly where to apply the Heisenberg Cut, but Heisenberg would probably be surprised to learn that by 2018 physicists would have confused themselves so much over quantum mechanics that they came to believe single photons are observers. 
What the Frauchinger-Renner paradox therefore establishes is that quantum mechanics can result in inconsistent predictions so long as we do not add a definition for what a measurement device is to the axioms of quantum mechanics.

This problem could easily be remedied---after all, we would just need to write down a definition. However, the definition for a measurement of course should not just remove the risk of inconsistent predictions but also agree with observations, and this just returns us to the question of where to apply the cut. 

It was argued in \cite{relano2018decoherence,zukowski2021physics} that the Frauchinger-Renner paradox can be resolved by taking into account decoherence. But decoherence is still a unitary and linear process that is described by the Schr\"odinger equation. It can therefore not give rise to the measurement update, so this still has to be added to the axioms. 

Decoherence can to some extent be used to identify the circumstances under which the measurement update should be applied, but this idea has its problems too. We will comment on this in more detail in Section \ref{ssec:dec}. For now, let us just note that decoherence alone simply will not evolve a system into a single detector eigenstate, and hence does not agree with what we observe. Tracing out the environment gives us a mixed state, but that is still not what we observe, not to mention that taking this trace is not a physical process, and therefore doesn't change anything about the state of the system.  
 
 \subsection{The Classical Limit}
 \label{ssec:classical}
 
Before quantum mechanics, there was classical mechanics, and classical mechanics still describes most of our observations correctly. Unfortunately, quantum mechanics doesn't correctly reproduce it.

It has long been known that recovering the classical time-evolution for suitably defined expectation values in quantum mechanics works properly only for integrable systems \cite{berman1978condition,zaslavsky1981stochasticity}. For chaotic systems, on the other hand, the quantum-classical correspondence breaks down after a finite amount of time \cite{combescure1997semiclassical,bambusi1999long}. As pointed out by Zurek \cite{zurek2003decoherence} (see also \cite{berry2001chaos}), this time may be long, but not so long that we can't observe it. Zurek estimates that the chaotic motion of Hyperion (a moon of Saturn) would last less than 20 years if we used the Schr\"odinger evolution for its constituents. Alas, it has lasted hundreds of millions of years.

Again, decoherence allegedly solves the problem. If one includes the interaction of Hyperion with dust and photons in its environment, then one sees that the moon becomes entangled with its environment much faster than its motion could significantly deviate from the classical limit. 

However, we have to note again that tracing out the environment is not a physical process. Therefore, all entanglement gives us is a very big entangled state. What we would have to do to get a classical non-linear motion of a localised object is to actually include the Collapse Postulate into the dynamical law. This shouldn't be so surprising: the non-linearity has to come from somewhere. 

To do this, we would have to know when and how the collapse happens, but we don't. Do the photons detect the moon? Or does the moon detect the photons? If there's neither photons nor dust, does the moon detect itself? And if the state collapses, then just exactly when do we update what part of the moon's wave function? These are not philosophical questions; these are questions about how to apply the axioms of our quantum machinery, and they are questions that we simply do not have an answer to. 

Another way to look at this problem was summarised by Klein \cite{klein2011limit}: the $\hbar \to 0$ limit of quantum mechanics just does not reproduce classical mechanics, unless one restricts oneself to special states (generalised coherent states) and specific types of potentials.  
 
 \subsection{Locality and Causality}
 \label{ssec:Loc}
 
The trouble with Hyperion brings us to the next problem. The collapse of the wave function in quantum mechanics is instantaneous; it happens at the same time everywhere in space. This ``spooky action at a distance'' \cite{Einstein1935EPR} understandably worried Einstein because it seems incompatible with the speed-of-light limit. We know now \cite{ghirardi1980general} that no information can be exchanged with the collapse of the wave function, but this doesn't explain how to apply the collapse postulate.

Consequently, people have debated for decades how to make the collapse compatible with relativistic invariance, and whether it requires backwards causation \cite{Percival1998Relativity,marolf2002relativistic,myrvold2002peaceful}. No resolution has been reached. 

We acknowledge, however, that the non-locality of the collapse is not a problem for the instrumentalist because, in the Copenhagen Interpretation, collapse is not necessarily a physical process, and is not related to any observable. So, in just which reference frame it happens does not matter; there are no predictions tied to this frame anyway. 

The reason we mention locality and causality is that these matter when we cross from special to general relativity, as we discuss next.

\subsection{Conservation Laws}
\label{ssec:Cons}

In Einstein's field equations
\begin{equation}
R_{\mu\nu} - \frac{1}{2} R g_{\mu \nu} + \Lambda g_{\mu \nu} = T_{\mu \nu}~,
\end{equation}
where the entries of the stress-energy tensor, $T_{\mu\nu}$ are $\mathbb{C}$-valued functions. In quantum field theory, however, the stress-energy tensor is operator-valued, $\hat T_{\mu \nu}$. 
To insert the stress-energy tensor from quantum field theory into Einstein's field equations, one thus has to give quantum properties to the left side of the field equations, which would require us to develop a theory of quantum gravity. Attempts to develop such a theory have been made since the 1950s, but have remained unsuccessful.

Another option is to instead convert the operator-valued stress-energy tensor into a $\mathbb{C}$-function. The most obvious way of doing this is to take its expectation value, $\langle \Psi| \hat T_{\mu \nu} |\Psi \rangle$, with respect to some quantum state $|\Psi\rangle$ that suitably describes the particle content of space-time. This is often referred to as semi-classical gravity. 

Most important for our purposes is that the semi-classical approximation is generally believed to be at least approximately correct in the weak-field limit and if fluctuations of the stress-energy tensor are small \cite{ford1982gravitational,Kuo1993SemiclassGrav}.

But the expectation value in the stress-energy tensor generically has to be updated upon measurement with the collapse of the wave function. And since this update is non-local, it violates the (contracted) Bianchi-identities which the left side of Einstein's field equations do fulfil and that are usually associated with local stress-energy conservation \cite{weinberg1972gravitation,Misner1973gravitation}.

Take for example a photon ($\gamma$) which passes through a beam splitter ($S$). Due to momentum conservation, this creates an entangled state 
\begin{eqnarray}
\frac{1}{\sqrt{2}}\left(|\gamma(p)\rangle |S(-p) \rangle + |\gamma(-p)\rangle |S(p)\rangle \right)~,
\end{eqnarray}
where $p$ is the momentum and we will assume that it is really the mean value of a suitably localised wave-packet. The expectation value of this entangled state is a sum of localised momentum distributions going out in opposite directions from the beam splitter, while the mean momentum transferred to the beam splitter is zero.

According to the axioms of the quantum machine, when we measure the photon we have to update the wave function. In this moment, the momentum of one detector suddenly increases by $p$ ($-p$) and that of the beam splitter switches to $-p$ ($p$). The momentum is conserved, but how did it get to the detector? Without general relativity there is no observable tied to this question, but in general relativity there is: even though it is too small to measure, something must have happened with the space-time curvature. But what?

Again this is a problem for which we simply do not have an answer. We do not have the mathematics to describe what happens with the gravitational field of a particle if its wave function collapses, not even approximately. 

However, this shows an issue with approaches to the measurement problem which claim that wavefunction collapse is just a nonlocal process (e.g. Shimony's `passion-at-a-distance' \cite{Shimony1993NoSig}): while they may work to some extent for non-relativistic many-particle quantum mechanics, they are difficult to reconcile with both relativistic quantum mechanics and gravity. And just letting go of relativistic covariance is not an option either since it is experimentally extremely well-confirmed  \cite{Mattingly2005Lorentz,Kosteleck2011LorentzData}. That is to say, while it cannot strictly speaking be ruled out that a nonlocal approach can be made ``local enough'' to agree with all available evidence, it seems like a stretch. The more pragmatic approach is to just look for an approach that is local and  relativistically covariant to begin with.

\section{What can we say about the solution?}

\subsection{Solution Requirements}

From the previous section, we see that a satisfactory solution of the measurement problem must achieve the following:

\begin{requirement}\label{R1}
  Agree with all existing data.
\end{requirement}

\begin{requirement}\label{R2}
  Reproduce quantum mechanics, including the Collapse Postulate (Axiom \ref{A4Collapse}) and Born's Rule (Axiom \ref{A5Born}), in a well-defined limit.
\end{requirement}

\begin{requirement}\label{R3}
  Give an unambiguous answer to the question of what a measurement device is, at least in principle.
\end{requirement}

\begin{requirement}\label{R4}
  Reproduce classical physics in a well-defined limit.
\end{requirement}

\begin{requirement}\label{R5}
  Resolve the inconsistency between the non-local measurement collapse and local stress-energy conservation.
\end{requirement}

Requirement 1 must be fulfilled by any scientifically adequate theory and we just add it for completeness. Requirement 2 recognises that within its domain, standard quantum mechanics is incredibly well-supported by data and a new theory would not become accepted without reproducing the achievements of its predecessor.  Requirement 3 is necessary to resolve the problem laid out in {\ref{ssec:Heisenberg}}, Requirement 4 the problem laid out in {\ref{ssec:classical}}, and Requirement 5 the problem discussed in {\ref{ssec:Loc}} and {\ref{ssec:Cons}}.

Requirements \ref{R2} and \ref{R4} have the same form---they both consist of ensuring the solution reproduces an established, well-proven theory in some well-defined limit. Their resolution might well be related; however, this does not have to be so, which is why we list them separately. Likewise, we would expect that requirements \ref{R2} and \ref{R4} can be used to show that requirement \ref{R1} is fulfilled. Again, however, this does not have to be the case---a limit might be well-defined and yet its result might just be in conflict with observations---so we list them separately. 

We added the phrase ``at least in principle'' in Req. \ref{R3} to make clear that no one expects it to be of much practical use to calculate from first principles what the arrangements of elementary particles in a detector are. To find out what a detector is, it is much more practical to just test whether it actually detects the thing we want to detect. However, even though it may be unpractical or even unfeasible to perform an exact calculation, we should be able to identify some general properties for what it takes for a collection of particles to act as a measurement device.

As mentioned previously, Req. \ref{R5} might be resolved by a theory of quantum gravity. However, the currently most well-developed approaches to quantum gravity do not address the measurement process. Similarly, these requirements could feasibly be met by a theory which isn't also a theory of quantum gravity. Therefore, we emphasise a theory of quantum gravity is neither necessary nor sufficient to solve the measurement problem.

\subsection{Solution Properties}

So what does it take to meet these requirements? Req. \ref{R5} requires a local evolution law that connects the initial state of the system with the observed measurement outcome. This means that either
 
\begin{soln}\label{S1}
  The wave function itself evolves according to Axiom \ref{A3} and is updated according to Axiom \ref{A4Collapse}, but it is an incomplete description. The physical state of the system is described by something else that evolves locally though not necessarily deterministically.
\end{soln}

or

\begin{soln}\label{S2}
  The wave function is the complete description of the system, but does not evolve by Axiom \ref{A3} and Axiom \ref{A4Collapse}. It evolves instead according to a different, local evolution law, that is necessarily non-deterministic.
\end{soln}

or

\begin{soln}\label{S3}
  A combination of Solutions \ref{S1} and \ref{S2}.
\end{soln}

Solution \ref{S1} is what is commonly called a ``hidden variables'' theory. We will instead adopt the convention of \cite{Maudlin1995ThreeMP} and refer to it as an ``additional variables'' theory, to acknowledge that the variables may not, in fact, be hidden---the variables are merely not included in axioms \ref{A1}-\ref{A6}. It should be noted that according to this definition, details of the detector and environment (as they appear in the decoherence approach) count as additional variables. We will come back to this point later. 

For the instrumentalist, all three solutions come down to local, deterministic evolution laws with additional variables. We here use the definition for deterministic from {\cite{cavalcanti2012bell}}, which is that measurements have definite outcomes (probability 0 or 1). Using this terminology, the laws may be non-predictable, despite being deterministic.  

This can be seen by noting that any local, non-deterministic evolution law can be rewritten into a local, deterministic evolution in a theory with additional variables: any time the evolution law is indeterministic (which could be continuously), we encode the possible time-evolutions with an additional variable. 

To give a concrete albeit trivial example: we can make the Collapse Postulate deterministic by just using the eigenstate that is the outcome of the collapse as an additional variable. The collapse is then ``determined'' by the ``additional variable''. 
Of course, this is somewhat pointless, because in quantum mechanics this additional variable is unpredictable (hence would well deserve the name ``hidden''), but it serves to show that additional variables \emph{could} be used to describe the process.

The difference between the solutions is then merely what we call the ``wave function''. Do we reserve the term for that which evolves under Schr\"odinger equation (Solution \ref{S1}) which has to come out of the mathematics at least in some limit (Req. \ref{R2}), or are we okay with using the term for whatever it is that we use to describe the state of the system (Solutions \ref{S2} and \ref{S3})? Do we require wave functions to be elements of a Hilbert-space, regardless of their evolution law? Do we want them to factorise for separable systems? Do we want them to be green, or married, or all of the above? In the end, this is just a matter of definition. The instrumentalist doesn't care and concludes that all possible solutions to the problem can be covered with a local and deterministic evolution in a theory with additional variables. 

This, together with the undeniable resemblance between the von Neumann-Dirac equation and the Liouville equation, makes it plausible that quantum mechanics is indeed an ensemble description of an underlying statistical theory with the additional variables \cite{Hance2021Ensemble}. 

\subsection{Solution Parameterisation}
\label{ssec:para}

That this new underlying theory must explain just where the Heisenberg Cut is (Req. \ref{R3}) means that it has to bring in some new parameter to quantify how good the statistical approximation is. This parameter cannot be derived from quantum mechanics itself; it must be extracted from experiment. 

It is clear that the new transition parameter cannot be something as simple as just the number of particles, not least because that quantity is not in general well-defined. (How many particles are inside an atom? How many particles does the vacuum contain?) Seeing the problems of the decoherence approach (see \ref{ssec:dec}) it is also unlikely to be any measure of decoherence or entanglement, though the frequency or strength of interactions must play some role. Parameters based on the total mass or energy or gravitational self-energy have no \emph{a priori} relation to generic measurements, and parameters based on the classical limit (possibly using the action as a quantifier) create a circular problem, because we would need them to know how the classical limit works to begin with.

However, we have reason to be optimistic that this problem is solvable because we have observational limits on the Heisenberg Cut both from above and below, and current technologies are pushing both these limits: by bringing larger objects into quantum states, while at the same time shrinking the size of detectors. It is only a matter of time until experiment will reach a regime in which deviations from quantum mechanics become noticeable. However, this process could be accelerated if we knew better what to look for, which is a task for theory development. 

\section{Solution Attempts}

\subsection{Second Quantization}

 The problems we discussed in the previous sections do not disappear with second quantization. All quantum field theories are built on the basic axioms of quantum mechanics which we listed in Section \ref{sec:ax}. In quantum field theory we merely have more complicated ways of describing interactions, and calculating the time-evolution of the system and observables related to it. If anything, it has been argued these problems become more complicated in quantum field theory \cite{grimmer2022pragmatic}.
 
 \subsection{Understanding Quantum Mechanics}
 
 Some readers may wonder if it is possible that these problems will one day be solved within the context of quantum mechanics itself. Maybe the problem with the classical limit is just that no one has found the right limit. Indeed, it is widely known that generalised coherent states make a promising basis for taking the classical limit \cite{yaffe1982large,landsman2006between}. But, in light of the problems pointed out by Klein \cite{klein2011limit}, this would at least entail adding further axioms about what to do for obtaining the classical limit. Within the context of this present argument we would therefore have to consider it a new theory, because its set of axioms would not be equivalent to the one we listed earlier.
 
 \subsection{Wave Functions as Epistemic States}
 \label{ssec:epis}

One common strategy to explain away the measurement problem is to argue that the wave function is not an ontic but an epistemic state, and that its collapse is not a physical process. The collapse, so the argument goes, is merely an update of our knowledge about the system, and its non-locality therefore should not worry us. After all, as Bell put it, when the Queen dies, Prince Charles will instantaneously become King, yet no information had to be sent non-locally for this update \cite{bell2004lanouvellecuisine}. 

A common strategy to counter this argument is to point out that if the wave function is the complete description of the system, then there is nothing else the wave function could describe knowledge about \cite{Maudlin1995ThreeMP}. Therefore, most wave function-epistemic views require the wavefunction to be an incomplete representation of the system.

While attempts have been made at formalising this view \cite{Spekkens2007EpistTot,Harrigan2010OMF} and using it to come to no-go theorems on the wavefunction being in some way epistemic \cite{Pusey2012Reality,Patra2013NoGo,Ruebeck2020Epistemic}, these formalisations have fundamental issues \cite{Schlosshauer2012Implications,Oldofredi2020Classification,Hance2021Wavefunctions}. Therefore, these no-go theorems contribute little to telling us whether the wavefunction can or cannot be epistemic.

In reality, the two sides of this argument haven't much advanced since the debate between Einstein and Bohr, and at this point it seems unlike they ever will. Let us therefore note that the problems listed in Section \ref{sec:prob} exist regardless of whether one believes the wave function is epistemic or ontic or how one wants to interpret the collapse. Quantum mechanics is unsatisfactory for the instrumentalist simply because we cannot answer questions about physical reality with it: just what properties does a measurement device need to have? Just what happens with space-time curvature when a photon passes through a beam splitter? Saying that the wave function is epistemic doesn't answer these questions. 

\subsection{Decoherence}
\label{ssec:dec}

The virtues of the decoherence program can be briefly summarised as follows. Given any system that includes a prepared state, detector, and environment, a detector is a subsystem that can keep a record of at least one aspect of another subsystem, which is the prepared state one wants to measure. To be able to keep a record of (some property of) the prepared state, the detector itself must have states that are stable under interaction with the rest of the system, which is the environment. 

These stable detector states are often called `pointer states' and we will denote them with $|I\rangle$. They keep a record of the prepared state's projection on the eigenstate corresponding to the pointer state, i.e., one gets a product state $|I\rangle |O_I\rangle$. Any superposition of pointer states would rapidly decohere under interaction of the environment, hence not keep a record of what we are interested in.

To describe the process of decoherence formally, one takes the density matrix of pointer states, prepared states, and environment. One estimates how quickly they become entangled and how much this affects the relative phases between the pointer states. This is the process of decoherence. It must be stressed that this process is fully described by the Schr\"odinger equation. After that, one traces out the environment, and obtains a mixed state whose probabilities are given by Born's rule. 

What one learns from this is that a useful detector, loosely speaking, must be large enough so that superpositions of its pointer states rapidly decohere. Decoherence hence gives us a criterion for identifying detector pointer states by what Zurek termed `einselection' by the environment \cite{zurek2003decoherence}.

But we do not observe mixed pointer states any more than we observe their superpositions. We only observe detectors in pointer states. In terms of the density matrix, we observe a matrix that has one entry equalling one, somewhere on the diagonal, and all other entries equal zero. The result of decoherence, however, is a density matrix with the Born probabilities on the diagonal.

Thus, while decoherence explains why we do not measure cats that are in a superposition of dead and alive, it does not explain why we do not measure cats that are 50\% dead and 50\% alive (a classical mixture) either \cite{Schlosshauer2004Decoherence}. 
To agree with observations, the wave function, or its density matrix, respectively, must therefore still be updated upon measurement. 

Another way to see that decoherence does not solve the measurement problem is noting that it is based on counterfactual reasoning: the typical initial state of the system will, under the Schr\"odinger equation, evolve into a final state that is highly entangled with the environment, susceptible to decoherence, and hence not what we observe. According to the decoherence program, the state we observe is instead an (almost) decoherence-free subspace which is exactly what we generically do \emph{not} expect. But the decoherence program gives us no clue as to how we get from evaluating the amount of decoherence in a state we do not observe to the not-decohering state we do observe. 

This discrepancy raises the question of whether the notion of entropy we use in quantum mechanics, that increases under an evolution that we do not actually observe, can possibly be correct. Indeed, if the wave function describes an ensemble average, we do not expect a notion of entropy derived from it to be meaningful. 

The decoherence program suffers from another problem, as pointed out by Kastner \cite{Kastner2014Einselection}. It requires one to specify a division between the observed system, the detector, and the environment already. Without that division, one does not know what the environment is that one should trace out. For this reason, decoherence does not allow us to define what a detector is. It merely quantifies certain properties that we know detectors do have. 

We do not mean to deny the usefulness of studying and quantifying decoherence and entanglement. But they can ultimately not solve the problem of how to define a measurement, because these properties are basis-dependent. They will just reformulate the question into one about the choice of basis or the division into subsystems, respectively.

\subsection{Many Worlds} 

The many-worlds interpretation \cite{everett1957relative,dewitt2015many} and similar approaches are often claimed to be simpler than the Copenhagen Interpretation \cite{Faye2019SEPCopenhagen} because they do not require the Collapse Postulate. The fact that we only observe detector eigenstates is allegedly explained by the branching of the wave function and is supposedly a consequence of the Schr\"odinger equation alone.

Alas, this is just not the case. To make a prediction for a measurement outcome in a many-words approach, one has to replace the Collapse Postulate with sufficiently many assumptions that achieve the same. Those are normally stated as assumptions about what constitutes an observer or a detector or a branching event \cite{Sebens2018OriginsEverettCarroll}; in any case, they are clearly not any simpler than the Collapse Postulate. 

The easiest way to see that many worlds does not do away with the Collapse Postulate is to note that if it was possible to make predictions for our observations using the Schr\"odinger equation alone, then this would be possible in any interpretation of the mathematics. One therefore clearly needs the Collapse Postulate or at least equivalent assumptions in many worlds, regardless of how they are called or interpreted. 
 
This is not to say that many world approaches are wrong. From the instrumentalist perspective, they are as good or as bad as the Copenhagen Interpretation. Anybody who doubts this statement is strongly encouraged to make a prediction with the many worlds interpretation and try to figure out how this differs from one made with the Copenhagen Interpretation. 
 
\subsection{Bohmian Mechanics}

Bohmian Mechanics \cite{Bohm1952Bohm1,Bohm1952Bohm2} comes in two different versions, one in which the equilibrium hypothesis is counted as an axiom, and one in which it is not an axiom, but merely approximately fulfilled in the situations we typically observe in the laboratory.

Bohmian Mechanics with the equilibrium hypothesis is mathematically equivalent to the Copenhagen Interpretation in the sense that one can be derived from the other. They make exactly the same predictions. Bohmian Mechanics is usually formulated in position space, but one can easily extend this definition just by requiring it to respect invariance under basis transformations.

Since Bohmian Mechanics with the equilibrium hypothesis is equivalent to the Copenhagen Interpretation, it cannot solve the problems laid out in Section \ref{sec:prob}. It adds rather than removes non-locality, does not give us any clue about how to define a detector, and doesn't help us take a classical limit. 

The reason Bohmian Mechanics is often said to solve the measurement problem is that the outcome of the time-evolution, interpreted suitably, is a detector eigenstate. In Bohmian Mechanics, one has a distribution of particles but interprets the actual ontic state of the system to be only one of them. Loosely speaking, Bohmian Mechanics combines the Schr\"odinger evolution and the Collapse Postulate to one local evolution for the particle and a non-local one for the guiding field. Since by assumption there is only one particle in the initial distribution, there is only one final outcome.\footnote{Though this leaves one with the long-standing problem of the empty waves.}

This solution however only works if one measures positions, so if one wants this solution to go through one has to argue that the only thing we ever measure are really positions of particles and everything else is derived from that. Given that the Collapse Postulate also brings the system into a detector eigenstate, one thus doesn't gain any advantage from switching to Bohmian Mechanics, one just gets this new headache. 
Further, since the ontology of Bohmian mechanics is itself non-local, it makes it even more difficult to conceive of a solution to the measurement problem. 

Again, this is not to say that Bohmian Mechanics is wrong. Being equivalent to the Copenhagen Interpretation, it is isn't any better or worse: Nikolic proposed comparing Bohmian Mechanics to the Coulomb gauge of electrodynamics \cite{nikolic2006many}. It seems non-local, and though it doesn't give rise to non-local observables, the explicit non-locality makes it difficult to generalise the formulation. It is quite possibly for this reason that quantum field theories based on Bohmian mechanics have been complete non-starters \cite{durr2004bohmian,durr2014can}. See also \cite{wallace2022sky} for more about the difficulty of generalising different interpretations of quantum mechanics to quantum field theory. Bohmian Mechanics may suffer from more severe problems than its failure to solve the measurement problem (see e.g. \mbox{\cite{Einstein1953,einstein2011elementary,helling2019no}}), but we will not discuss these here because it is not relevant for our purposes.

Bohmian Mechanics without the equilibrium hypothesis \cite{Valentini1991Nonequilib1,Valentini1991Nonequilib2} is distinct from quantum mechanics, and to our best knowledge it has not helped solving the measurement problem. Since it has to reproduce the equilibrium hypothesis in the situations we typically encounter in the laboratory, it is also implausible that it would be of use.

\subsection{Other Interpretations} 
At this point, it should be clear that the problem can't be solved by reinterpreting the maths: we actually need new maths. If we cannot derive what a measurement device is, or what the source of gravity is in one interpretation, we can't do it in any interpretation. This means that QBism \cite{fuchs2010qbism,Fuchs2014QBism,fuchs2017notwithstanding}, the modal interpretation \cite{dieks1998modal}, the previously mentioned statistical interpretation \cite{ballentine1970statistical}, the transactional interpretation \cite{cramer1986transactional,kastner2013transactional}, Rovelli's relational interpretation \cite{rovelli1996relational,adlam2022information}, or Smolin's ensemble interpretation \cite{smolin2012real}, or any other reinterpretation of the mathematics may in the best case make us feel better about quantum mechanics, but they can't actually solve the problems we laid out in Section \ref{sec:prob}. Those aren't just questions whose answers are difficult to calculate; they're questions whose answers can't be calculated in quantum mechanics---regardless of its interpretation. 

We want to stress however that we certainly do not mean to say that it is useless to think about different interpretations of quantum mechanics. This is because some interpretations may make it easier to answer certain questions. A good example is the question of arrival time, a quantity that is notoriously difficult to calculate in the Copenhagen Interpretation, but that was recently successfully calculated using Bohmian Mechanics \cite{das2019arrival}.  

\subsection{Collapse Models}

Collapse Models (be they gravitational, such as the Penrose-Diosi model, or spontaneous \cite{bassi2003dynamical}, such as the GRW \cite{Ghirardi1986GRW} and CSL \cite{ghirardi1990csl} models) have a chance to actually solve the problem, because they are not just reinterpreting the same mathematics. In accordance with what we discussed in \ref{ssec:para}, they all bring in new parameters that quantify the deviation from standard quantum mechanics. However, the currently existing collapse models run into a well-known problem: Bell's theorem. 

Remember that we can interpret any non-deterministic evolution as a deterministic evolution with additional hidden variables. This means, so long as collapse models fulfil the assumptions for Bell's theorem, they have to violate local causality (or they cannot reproduce observations). The currently used collapse models are therefore either still non-local \cite{Tumulka2006RelGRW}, or the evolution law explicitly contains the basis that the evolution collapses into. In the latter case they either violate statistical independence, or one is forced to assume that measurements can only be made in one particular basis (usually the position basis). 

\section{The role of statistical independence}
\label{sec:SI}

We have argued above that quantum mechanics suffers from several problems, and that any solution to the problem can be expressed as a local, deterministic theory with additional variables. 

However, we can measure different observables of the same prepared state, and different observables correspond to different detector pointer states. If the evolution into the detector eigenstate is to be local, then the additional variables which determine the outcome must contain information about the pointer states from the outset. If the prepared state only gets this information by the time it arrives at the detector, then the collapse will generically have to be faster than light---this is exactly what happens in standard quantum mechanics. 

If we want to avoid this, then the additional variables, commonly denoted $\lambda$, must be correlated with the measurement settings. This is known as a violation of statistical independence, or a violation of measurement independence. If $\rho$ is the probability distribution of the hidden variables, and $X$ are the detector settings (of possibly multiple detectors), then a violation of statistical independence means $\rho(\lambda|X) \neq \rho(\lambda)$. Theories with this property have been dubbed superdeterministic \cite{Hossenfelder2020Rethinking,Hossenfelder2020Perplexed}. 

To our best knowledge we do not presently have any theory that fulfills requirements \ref{R1}-\ref{R5}, but from our above arguments we do know that any such theory has to violate statistical independence. Seeing that most of what we argued above (except the reference to Frauchinger-Renner) could have been said 50 years ago, it is curious that progress on this has been so slow. We would like to offer some thoughts on why that may be. 

One reason we don't yet have a solution to the problems of Section \ref{sec:prob} is almost certainly that so far it just wasn't necessary. Quantum mechanics in its present form has worked well, made a stunning amount of correct predictions, and for the most part saying that we know a detector when we see one is sufficient to make predictions with quantum mechanics' mathematical machine. 

Let us not forget that it wasn't until 1982 \cite{Aspect1982BellTest} that violations of Bell's inequality were conclusively measured. Until about a decade ago, interpretations of quantum mechanics were discussed primarily by philosophers, simply because they weren't relevant for physicists. The re-approach between philosophy and physics in quantum foundations is a quite recent development, and it has been driven by technological advances. 

Even now, the problems that we outlined in Section \ref{sec:prob} have grown of interest to the instrumentalist, but not yet for the experimentalist. This is about to change though. Soon, experiments will probe into the mesoscopic range, and investigate the question of just when a device ceases to be a useful detector. Further, tests of the weak-field limit of quantum gravity are on the way \cite{Bose2017EntWitnessQG,Marletto2017GravIndEnt}---granted, the latter experiments weren't designed to test the Collapse Postulate, but they examine the parameter range where the questions raised above are also relevant.

Another likely reason why we haven't yet managed to solve the measurement problems is that the only viable option---violating statistical independence---was discarded 50 years ago on purely philosophical grounds. For peculiar reasons, statistical independence became referred to as the `free will' or `free choice' assumption, which has discouraged physicists from considering that this assumption may just not be fulfilled, and hence that a local description of the measurement process may still be possible. It has already been explained elsewhere \cite{Hossenfelder2020Rethinking,Hossenfelder2020Perplexed} that this terminology is meaningless; whether statistical independence is violated or not bears no relevance for the existence of free will.
The option of violating statistical independence was so strongly discouraged that as of today many physicists do not even know that Bell's theorem does not generally rule out local and deterministic completions with additional variables \cite{Schlosshauer2013Attitudes}.

A third reason that probably added to the lack of interest in the option is that the additional variables are often interpreted as new degrees of freedom that reside inside elementary particles. This possibility is strongly constrained by experiments, and has only become more unappealing the more thoroughly we have tested the Standard Model of particle physics.

This, however, is based on a misunderstanding. The additional variables don't need to be new variables, and they don't need to reside on short distance scales; they merely need to be variables that don't appear in the standard axioms A1-A6. Further, as we mentioned above, the additional variables may merely be a stand-in for a non-deterministic (or non-computable) evolution law \cite{Hance2022Supermeasured}. 

Most importantly though, there is no reason why the additional variables must be located inside particles or located anywhere for that matter.
They could for example be the details of the detector or the environment or more generally variables that quantify large-scale properties or correlations. This means the solution we seek for may not to be found on the route of ontological reductionism that has preoccupied thinking in the foundations of physics for the past century (i.e., building bigger particle colliders won't solve this problem).

Indeed, as we have argued in \cite{Hance2021Ensemble}, the additional variables in Bell's theorem are better interpreted as labels for trajectories (which also explains why they can alternatively be understood as encoding a non-deterministic evolution law). 
Whatever the reason for the slow progress, we think that the problems we have laid out here are eminently solvable with existing mathematics and will become accessible for experimental test in the near future. 

\section{What is it good for?}

The brief answer to the question of what solving the measurement problem is good for (besides solving the problem, that is), is that we don't know. However, we can make some speculations.

For one thing, it might be that the underlying theory which solves the measurement problem turns out to be deterministic again, and explains the seeming indeterminism of quantum mechanics as being epistemic in origin. In this case, it stands to reason that the theory would allow us to overcome limits and bounds set by quantum mechanics on measurement accuracy with suitably configured experiments. 
This could turn out to be useful for many things, not least for quantum metrology and quantum computing. 

However, it could turn out the go the other way. If deviations from the Schr\"odinger equation become important for, say, quantum computers beyond a certain number of qubits (as they are in Palmer's Invariant Set Theory \cite{Palmer2020Discretization}), then maybe large quantum computers will be impossible \cite{slagle2021testing,Hance2021ExpIST}.  

More generally, understanding in which cases a measurement process occurs and just what happens would almost certainly improve our ability to control quantum states.

That is to say, solving the measurement problem is not just a philosophical enterprise. Its solution will quite possibly have technological applications. 

\section{Summary}

We have identified several different aspects of the measurement problem in quantum mechanics, and argued that to solve these problems we need a new theory. Deviations from quantum mechanics will likely become experimentally accessible in the near future. However, a theoretical understanding of the measurement process could greatly speed up the discovery. In a modest attempt to contribute to progress on the matter, we have listed five requirements that any satisfactory solution of the measurement problem must fulfil. 

\bigskip

\textit{Acknowledgements:} We thank Emily Adlam, Tim Palmer, Hrvoje Nikolic, and Ken Wharton for helpful discussions and feedback.
JRH acknowledges support from the University of York's EPSRC DTP grant EP/R513386/1, and the EPSRC Quantum Communications Hub (funded by the EPSRC grants EP/M013472/1 and EP/T001011/1). SH acknowledges support by the Deutsche Forschungsgemeinschaft (DFG, German Research Foundation) under grant number HO 2601/8-1.  




\bibliographystyle{unsrturl}
\bibliography{ref.bib}

\end{document}